\documentclass{article}
\usepackage[utf8]{inputenc}
\usepackage{amssymb,amsmath,amsfonts}
\usepackage[all]{xy}
\usepackage{graphicx}
\usepackage{amsmath}
\usepackage{amsthm}
\usepackage{amsfonts}
\usepackage{amssymb}%
\usepackage{xcolor}
\providecommand{\U}[1]{\protect\rule{.1in}{.1in}}
\newtheorem{theorem}{Theorem}

\newtheorem{idea memo}[theorem]{Idea Memo}

\newtheorem{notation}[theorem]{Notation}

\newtheorem{Definition}[theorem]{Definition}
\newtheorem{Theorem}[theorem]{Theorem}

\mathchardef\mhyphen="2D

\title{Quantum Fields as Category Algebras}
\author{Hayato Saigo\footnote{h\_saigoh@nagahama-i-bio.ac.jp}\\
Nagahama Institute of Bio-Science and Technology}

\date{11th September 2021}

\begin{document}

\maketitle

\begin{abstract}
In the present paper we propose a new approach to quantum fields in terms of category algebras and states on categories. We define quantum fields and their states as category algebras and states on causal categories with partial involution structures. By utilizing category algebras and states on categories instead of simply considering categories, we can directly integrate relativity as a category theoretic structure and quantumness as a noncommutative probabilistic structure. Conceptual relationships with conventional approaches to quantum fields, including Algebraic Quantum Field Theory (AQFT) and Topological Quantum Field Theory (TQFT), are also discussed.
\end{abstract}

\section{Introduction}

Quantum fields are the most fundamental entities in modern physics. Intuitively, the notion of quantum field is the unification of relativity theory and quantum theory. However, the existence of a nontrivial interacting quantum field model defined on a four-dimensional Minkowski spacetime, which is covariant with respect to the Poincar\'{e} group, has not yet been proven. In axiomatic approaches to quantum field theory, there have been shown many fundamental theorems including no-go theorems such as Haag's theorem \cite{HaagTheorem,GeneralizedHaagTheorem} which implies that "interaction picture exists only if there is no interaction" \cite{SW} through the clarification of the concept of quantum field (see \cite{SW,HAG} and references therein).
To put it roughly, we cannot go beyond the free fields if we stick to the axioms that we take for granted in conventional quantum field theories. 

In this paper, we propose a new approach to quantum fields:
The core idea is to deal with quantum fields in terms of category algebra, which is noncommutative in general, over a rig ("ring without negatives"), i.e., algebraic system equipped with addition and multiplication, where the category and the rig correspond to the "relativity" aspect and the "quantum" aspect of nature, respectively. By utilizing category algebra and states on categories instead of simply considering categories, we can directly integrate relativity as a category theoretic structure and quantum nature as a noncommutative probabilistic structure. The cases where the rig is an algebra over $\mathbb{C}$, the field of complex numbers, are especially important for the our approach to quantum fields. For other regions of physics such as classical variational contexts, the tropical semiring (originally introduced in \cite{SIM}), i.e., a rig with "min" and "plus" as addition and multiplication, will be useful. The author believes that it is quite interesting to see the quantum-classical correspondence from the unified viewpoint from category algebras over rigs.

As is well known, the essence of the relativity is nothing but the structure of possible relationships between possible events. If we assume the structure of relationships between events, we can essentially reconstruct the relativity structure. More concretely, in \cite{MAL} it is shown that two future-and-past-distinguishing Lorentzian manifolds are conformally equivalent, if and only if the associated posets are isomorphic, where the poset consists of events and the order relation defined by the existence of future-directed causal curves, based on \cite{HKM}: What really matters are causal relationships (for details see \cite{GSC} and reference therein).
This viewpoint is quite essential and there is an interesting order-theoretic approach to spacetime (for example, "causal set" approach \cite{BLMS} ). However, to deal with the off-shell nature of quantum fields, which seems to be essential in modelling interacting fields on the spacetime, we need to take not only causal relationships but also more general relationships between spacelike events into consideration. Then, the question arises: how should we generalize a framework of previous approaches? 

The strategy we propose is to think a category $C$, which is a generalization of both of ordered sets (causality structures) and groups (symmetry structures), "as" relativity in a generalized sense. 
More concretely, we identify the notion of causal category  equipped with partial involution structure introduced in Section 2, as the generalized relativity structure. To combine this relativity structure with quantum theory, which can be modelled by noncommutative rigs, especially effectively by noncommutative algebras over $\mathbb{C}$ as history has shown, we need noncommutative algebras that reflect the structures of categories: Category algebras are just such algebras. As categories are generalized groups, category algebras are generalized group algebras.

The above discussion intuitively explains why we use category algebras to model quantum fields. For simplicity, in this paper we focus on a category algebra which satisfies a suitable finiteness condition. 
Importantly, the category algebras can be considered as generalized matrix algebras over R as well as generalized polynomial algebras \cite{SAI}, which provides a platform for concrete and flexible studies and calculations. The extension to a larger algebra is of course of interest, but the category algebras we focus already have rich structure as covariance and local structure of subalgebras reflecting the causal and partial involution structure of the category as we will see in Section 3. By focusing on these structures, we can also see the conceptual relationship between our approach and the preceding approaches such as Algebraic Quantum Field Theory (AQFT) \cite{HK,HAG} and Topological Quantum Field Theory (TQFT) \cite{ATI,WIT}.

Identifying a quantum field to be a category algebra over a rig, the next problem  which is treated in Section 4 is how to define a state of it. In general, the notion of state on $^{\ast}$-algebra over $\mathbb{C}$ is defined as a positive normalized linear functional. We can naturally extend the notion in the context of algebras with involution over rig (\cite{SAI} for details). We call the states on category algebras as states on categories. If the number of objects in category is finite, states can be characterized by functions on arrows satisfying certain conditions \cite{SAI}, which is a generalization of the result in \cite{CIM} for groupoids with finite number of objects. More generally, to define a state on category whose support is contained in a subcategory with finite numbers of objects is equivalent to define the corresponding function which assign the weight to each arrow. By considering such states, we can see a quantum mechanical system as an aspect of the quantum field. This viewpoint will shed a new light on the foundation of quantum theory. 

For the study of quantum fields, a localized notion of state, or a "local state" \cite {WER,OOS} is important. We can define the counterpart of the notion, originally studied in AQFT approach, as the system of states on certain subalgebras of category algebras called local algebras introduced in section 3. These matters will be explained in Section 4 with more clarification of the conceptual relationship with AQFT and TQFT. The discussion in Section 4 will provide a new basis for generalizing the DHR(Doplicher-Haag-Roberts)-DR(Doplicher-Roberts) sector theory\cite{DHR1,DHR2,DHR3, DHR4,DR1,DR2,DR3} and developing Ojima's micro-macro duality \cite{OJI1,OJI2} and quadrality scheme \cite{OJI3} from the viewpoint of category algebras and states on categories.

In the last section, we will discuss the prospect of research directions based on our framework. In addition to the importance of mathematical research, such as taking topological or differential structures into account, there is the challenge of integrating various approaches to quantum fields, and of conducting research on quantum foundations, based on our framework. These are where new concepts such as quantum walks on categories will be useful. One of the most exciting problems is, of course, to construct a model of a non-trivial quantum field with interactions. The author hopes that the present paper will be a small new step towards these big problems.

\section{Structure of Dynamics as Category}

In this section, the "relativistic structure" as the basic structure of dynamics, consisting of possible events and relations (or "processes") between them, is formulated in terms of category theory.




\subsection{Definition of Category}


A category is a mathematical system composed of entities called objects and arrows (or morphisms) satisfying the following four conditions.
\begin{quote}
\textbf{Condition 1.} For any arrow $f$ there exist an object called $\mathrm{dom}(f)$ and another object $\mathrm{cod}(f)$, which are called the domain of $f$ and the codomain of $f$, respectively. 
\end{quote}

When $\mathrm{dom}(f)=X$ and $\mathrm{cod}(f)=Y$ we denote as
\begin{align*}
f: X \longrightarrow Y
\end{align*}
or
\begin{align*}
X \xrightarrow[]{~f~} Y.
\end{align*}
Arrows are also denoted in any direction, not only from left to right as above. 

\begin{quote}
\textbf{Condition 2.} For any pair of morphism $f,g$ satisfying $\mathrm{dom}(g) = \mathrm{cod} (f)$
\begin{align*}
Z \xleftarrow[]{~g~} Y \xleftarrow[]{~f~} X,
\end{align*}
there exist an arrow $g\circ f$
\begin{align*}
Z \xleftarrow[]{~g \circ f~} X
\end{align*} 
called the composition of $f,g$.
\end{quote}

For the composition of arrows, we assume the following conditions:

\begin{quote}
\textbf{Condition 3. (Associative Law)}
For any triple $f,g,h$ of arrows satisfying $\mathrm{dom}(h)=\mathrm{cod}(g)$ and $\mathrm{dom}(g)=\mathrm{cod}(f)$,
\begin{align*}
(h \circ g) \circ f
= h \circ (g \circ f)
\end{align*}
holds.
\end{quote}




\begin{quote}
\textbf{Condition 4. (Identity Law)}
For any object $X$ there exists an arrow called \textbf{identity arrow} $1_X : X \longrightarrow X$. For any arrow $f: X \longrightarrow Y$ 
\begin{align*}
f \circ 1_X = f = 1_Y \circ f
\end{align*}
holds.
\end{quote}


By the correspondence from objects to their identity arrows, objects can be considered as special kinds of arrows, by identifying each object $X$ with its identity arrow $1_X$. 



To sum up, the definition of a category is as follows:

\begin{Definition}[Category]
A category is a system composed of two kinds of entities called objects and arrows, equipped with domain/codomain, composition and identity, satisfying associative law and identity law.  
\end{Definition}

In a category, we can define the "essential sameness" between objects via notion of invertible arrows (isomorophism):

\begin{Definition}[Invertible Arrow (Isomorphism)]
Let $\mathcal{C}$ be a category. An arrow $f:X\longrightarrow Y$ in $\mathcal{C}$ is said to be invertible in $\mathcal{C}$ if there exists some arrow $g:Y\longrightarrow X$ such that 
\[
g\circ f=1_X,\:\:f\circ g=1_Y.
\]
An invertible arrow in $\mathcal{C}$ is also called an isomorphism in $\mathcal{C}$. 
\end{Definition}

There are many categories whose collection of arrows is too large to be a set. In the present paper we focus on small categories:

\begin{Definition}[Small Category]
A category $\mathcal{C}$ is called small if the collection of arrows is a set. 
\end{Definition}

Let us see the examples of small categories which are used in the present paper.

\begin{Definition}[Preorder]
A pair $(P,\rightsquigarrow)$ of a set $P$ and a relation $\rightsquigarrow$ on $P$ satisfying
$p\rightsquigarrow p $ for any $p\in P$ and 
\[
p\rightsquigarrow q \;\text{and} \; q\rightsquigarrow r \Longrightarrow p\rightsquigarrow r 
\]
for any $p,q,r\in P$ called a preordered set. The relation $\rightsquigarrow$ on $P$ is called a preorder on $P$.
The preordered set $(P,\rightsquigarrow)$ can be viewed as a category whose objects are elements of $P$, when we define the relation $p\rightsquigarrow q$ between $p,q$ as the unique arrow from $p$ to $q$. Conversely, we can define a preordered set as a small category such that for any pair of objects $p,q$ there exists at most one arrow from $p$ to $q$. 
\end{Definition}

Note that the notion of preorder is a generalization of a partial order and an equivalence relation. 
As a special extreme case of the concept of preordered sets, we have the following:  

\begin{Definition}[Indiscrete Category and Discrete Category]
An indiscrete category is a small category such that for any pair of objects $C,C'$ there exists exactly one morphism from $C$ to $C'$. A discrete category is a small category such that all arrows are identity arrows.
\end{Definition}

Note that an indiscrete category corresponds to a complete graph and that any set can be considered as a discrete category.

On the other hand, the notion of a group, which is essential in the study of symmetry, can also be defined as a small category as follows: 

\begin{Definition}[Monoid and Group]
A small category with only one object is called a monoid. A monoid is called a group if all arrows are invertible.
\end{Definition}

To see the equivalence between the definition of the group as a category, just define the arrows in a group as a category as the elements in the group as a set, and the unique identity arrow (which can be identified with the unique object) as the identity element of the group. 

By definition, the concept of monoid is a generalization of that of a group, allowing the existence of non-invertible arrows. The concept of a groupoid is another generalization of that of a group:   

\begin{Definition}[Groupoid]
A small category is said to be a groupoid if all arrows are invertible.
\end{Definition}

As for the importance of groupoids in physics, see \cite{CIM} and references therein, for example. From the mathematical point of view, the present paper is based on an extension of the previous work \cite{CIM} on groupoid algebras over $\mathbb{C}$ into category algebras of an arbitrary (small) category over a (in general, noncommutative) rig $R$, i.e., "ring without negatives" (algebraic system with addition and multiplication), which will be introduced in the next section. Even in the case of $R=\mathbb{C}$ this extension physically means allowing irreversible processes, since a category can be seen as a generalized groupoid allowing invertible arrows in general. Involution structure of the category algebra is provided by partial involution structure of the category, as we will see in the next section ($^{\dagger}$-category introduced there can be seen a generalization of a groupoid).

     

A functor is defined as a structure-preserving correspondence between two categories:

\begin{Definition}[Functor (Covariant Functor)]
Let $\mathcal{C}$ and $\mathcal{C}'$ be categories.
A correspondence $F$ from $\mathcal{C}$ to $\mathcal{C}'$ which maps object and arrows in $\mathcal{C}$ to objects and arrows in $\mathcal{C}'$ is said to be a covariant functor, or simply a functor, from $\mathcal{C}$ to $\mathcal{C}'$ if it satisfies the following conditions:

\begin{enumerate}
\item It maps $f:X \longrightarrow Y$ in $\mathcal{C}$ to $F(f):F(X) \longrightarrow F(Y)$ in $\mathcal{C}'$.
\item $F(g \circ f) =F(g) \circ F(f)$ for any (compositable) pair of $f,g$ in $\mathcal{C}$. 
\item For each $X$ in $\mathcal{C}$, $F(1_X) = 1_{F(X)}$.
\end{enumerate}
\end{Definition}


\begin{Definition}[Contravariant Functor]

Let $\mathcal{C}$ and $\mathcal{C}'$ be categories.
A correspondence $F$ from $\mathcal{C}$ to $\mathcal{C}'$ which maps object and arrows in $\mathcal{C}$ to objects and arrows in $\mathcal{C}'$ is said to be a contravariant functor from $\mathcal{C}$ to $\mathcal{C}'$ if it satisfies the following conditions:

\begin{enumerate}
\item It maps $f:X \longrightarrow Y$ in $\mathcal{C}$ to $F(f):F(X) \longleftarrow F(Y)$ in $\mathcal{C}'$.
\item $F(g \circ f) =F(f) \circ F(g)$ for any (compositable) pair of $f,g$ in $\mathcal{C}$. 
\item For each $X$ in $\mathcal{C}$, $F(1_X) = 1_{F(X)}$.
\end{enumerate}
\end{Definition}

\begin{Definition}[Composition of Functors]
Let $F$ be a functor from $\mathcal{C}$ to $\mathcal{C}'$ and $G$ be a functor from $\mathcal{C}'$ to $\mathcal{C}''$. The composition functor $G\circ F$ is a functor from $\mathcal{C}$ to $\mathcal{C}''$ defined as $(G\circ F)(c)=G(F(c))$ for any arrow $c$ in $\mathcal{C}$.
\end{Definition}

\begin{Definition}[Identity Functor]
Let $\mathcal{C}$ be a category. A functor from $\mathcal{C}$ to $\mathcal{C}$ which maps any arrow to itself is called the identity functor. 
\end{Definition}

We can consider categories consisting of (certain kind of) categories as objects and (certain kind of) functors as arrows.

The concept of involution on category is important throughout the paper:

\begin{Definition}[Involution on Category]\label{involution on category}
Let $\mathcal{C}$ be a category. A covariant/contravariant endofunctor $(\cdot)^{\dagger}$ from $\mathcal{C}$ to $\mathcal{C}$ is said to be a covariant/contravariant involution on $C$ when $(\cdot)^{\dagger} \circ (\cdot)^{\dagger}$ is equal to the identity functor on $\mathcal{C}$. A category with contravariant involution which is identity on objects is called a $^{\dagger}$-category.
\end{Definition}

We conclude this subsection by defining the concept of natural transformation and related concepts. The concept of natural transformation can be seen as a generalization of the various concepts of transformations in mathematics and other sciences including physics.

\begin{Definition}[Natural Transformation]
Let $\mathcal{C}, \mathcal{D}$ categories and $F,G$ be functors from a category $\mathcal{C}$ to a category $\mathcal{D}$. A correspondence $t$ is said to be a natural transformation 
from $F$ to $G$
if it satisfies the following conditions:
\begin{enumerate}
\item $t$ maps each object $X$ in $\mathcal{C}$ to the corresponding arrow $t_X:F(X)  \longrightarrow G(X)$ in $\mathcal{D}$.
\item For any $f:X \longrightarrow Y$ in $\mathcal{C}$, 
\begin{align*}
t_Y \circ F(f) = G(f) \circ t_X.
\end{align*}
\end{enumerate}
The arrow $t_X$ is is called the $X$ component of $t$.

\end{Definition}


\begin{Definition}[Functor category]
Let $\mathcal{C}$ and $\mathcal{C}'$ be categories. The functor category $\mathcal{C'}^{\mathcal{C}}$ is a category consisting of functors 
from $\mathcal{C}$ to $\mathcal{C}'$ as objects and natural transformations as arrows (domain,codomain, composition, and identity are defined in a natural way). 
\end{Definition}

\begin{Definition}[Natural Equivalence]
An isomorphism in a functor category, i.e., an invertible natural transformation, is said to be a natural equivalence.
\end{Definition}



\begin{notation}
In the rest of the present paper, categories are always supposed to be small. The set of all arrows in a category $\mathcal{C}$ is also denoted as $\mathcal{C}$. $|\mathcal{C}|$ denotes the set of all objects, which are identified with corresponding identity arrows, in $\mathcal{C}$. 
\end{notation}

\subsection{Relativistic Structure as Category}

If we intuitively consider the spacetime degrees of freedom with the geometric notion of "set" of possible events, it is natural to think that the structure of dynamics of quantum can be modelled with "category" as the total system structure of relationships.

\begin{Definition}[Causal Category]
A category $\mathcal{C}$ equipped with a subcategory $\mathcal{C}^{cau}$ satisfying $|\mathcal{C}|=|\mathcal{C}^{cau}|$ is called a causal category. Arrows in $\mathcal{C}^{cau}$ are said to be causal.
\end{Definition}

Any category can be considered as a causal category by taking $\mathcal{C}=\mathcal{C}^{cau}$. Note that $|\mathcal{C}|$ is equipped with preorder $\rightsquigarrow$ defined as the existence of causal arrow between objects.

A typical example of causal categories is constructed as follows. For a spacetime (with inner degrees of freedom) $E$, usually modelled by a manifold and sometimes by a symmetric directed graph (as in lattice gauge theory \cite{WIL}), we can construct a category $\mathcal{C}=\mathcal{M}[E]$ whose objects and arrows are points and path between them. More precisely,
we consider $\mathcal{M}(E)$ as a subcategory of the "Moore path category" \cite{BRO} of $E$ consisting of smooth paths in the manifold case and as the free category of $E$ in the discrete case.
Then we can define $\mathcal{C}^{cau}$ as the subcategory consisting of "causal paths". For manifold case, the notion of causal paths can be defined as the paths whose tangent vectors are all in the future light cone. For the graph case, a path (i.e., an arrow in the free category) $c$ is said to be causal if $c=c'\circ c''$ implies $\mathrm{dom}(c')\rightsquigarrow cod(c')$ and  $\mathrm{dom}(c'')\rightsquigarrow cod(c'')$, where $\rightsquigarrow$ denotes a preorder previously defined on the set of vertices.

\begin{Definition}[Relevant Category]
Let $\mathcal{C}$ be a causal category and $\mathcal{O}$ be a subset of $|\mathcal{C}|$. The subcategory of $\mathcal{C}$ generated by
\begin{quote}
    arrows whose domain and codomain are in $\mathcal{O}$,\\
    causal arrows whose domain is in $\mathcal{O}$ and whose codomain is in $ |\mathcal{C}| \setminus \mathcal{O}$,\\
    causal arrows whose codomain is in $\mathcal{O}$ and whose domain is in $ |\mathcal{C}| \setminus \mathcal{O}$, and\\
    identity arrows (identified with objects) in $ |\mathcal{C}| \setminus \mathcal{O}$,
\end{quote}
is called the relevant category for $\mathcal{O}$ and denoted as $\mathcal{O}^{rel}$.
\end{Definition}


By the definition of relevant categories, the following structure theorem holds: 

\begin{Theorem}[Structure Theorem for Relevant Category]
Let $\mathcal{C}$ be a causal category and $\mathcal{O}$ be a subset of $|\mathcal{C}|$. Any arrow in the relevant category $\mathcal{O}^{rel}$ can be written in either of the following forms:
\[
c,c^{out}\circ c, c\circ c^{in},c^{out}\circ c \circ c^{in}, i,
\]
where $c$ denotes an arrow whose domain and codomain is in $\mathcal{O}$, $c^{out}$ denotes a causal arrow whose domain is in $\mathcal{O}$ and whose codomain is in $ |\mathcal{C}| \setminus \mathcal{O}$, $c^{in}$ denotes a causal arrow whose codomain is in $\mathcal{O}$ and whose domain is in $ |\mathcal{C}| \setminus \mathcal{O}$, and $i$ denotes an identity arrow in $|\mathcal{C}| \setminus \mathcal{O}$.

\end{Theorem}

The notion below is quite important to see the essence of relativistic structure.

\begin{Definition}[Spacelike separated]
Let $\mathcal{C}$ be a causal category and $\mathcal{O},\mathcal{O'}$ be a subset of $|\mathcal{C}|$. 
$\mathcal{O}$ and $\mathcal{O'}$ is said to be spacelike separated if there is no causal arrow between their objects.
\end{Definition}

By definition, two spacelike separated subsets are disjoint since identity arrows are causal. Moreover, we have the following directly from the structure theorem of relevant category: 

\begin{Theorem}[Nonexistence of Nontrivial Compositable Pair]\label{Structure of Causal Category}
Let $\mathcal{C}$ be a causal category and  $\mathcal{O},\mathcal{O'}$ be a pair of spacelike separated subsets of  $|\mathcal{C}|$. There is no pair of arrows $(c,c')\notin |\mathcal{C}|\times|\mathcal{C}|$ satisfying
$c\in \mathcal{O}^{rel}$, $c' \in \mathcal{(O')}^{rel}$ and 
$\mathrm{cod}(c)=\mathrm{dom}(c')$.
\end{Theorem}

For the application to quantum theory, the involution structure is important.
From now on, we consider a causal category with partial involution structure defined below:

\begin{Definition}[Partial Involution Structure on Category]
Let $\mathcal{C}$ be a category. A partial involution structure on $\mathcal{C}$ is a subcategory $\mathcal{C}^{\sim}$ equipped with an involution such that $|\mathcal{C}|=|\mathcal{C}^{\sim}|$. 

\end{Definition}

Note that any category $\mathcal{C}$ has trivial partial involution structure, since $\mathcal{C}$ is equipped with the involution structure $|\mathcal{C}|$ defined as $C^{\dagger}=C$.

The notion is important because the category $\mathcal{C}^{\sim}$ physically means the category consisting of "bidirectional" processes. Although this notion is a generalization of the core (i.e., the maximal groupoid in a category consisting of isomorphisms), it does not require the reversibility of the process in the meaning of invertible arrows as isomorphisms. The author believes that this generalization from groupoids to categories and from cores to partial involution structures is quite important for the application to physical phenomena which include irreversibility.

Based on the partial involution structure, we define the notion of relevant category with involution.

\begin{Definition}[Relevant Category with Involution]
    Let $\mathcal{C}$ be a causal category with partial involution structure $\mathcal{C}^{\sim}$. 
    The maximal subcategory $\mathcal{O}^{rel \sim}$ of $\mathcal{C}^{\sim}$ closed under the involution is called the relevant category with involution on $\mathcal{O}$.
\end{Definition}

The importance of the relevant categories with involution is that we can naturally define algebras with involution from them. We will see the details in the next section.


\section{Quantum Fields as Category Algebras}

In the previous section, we have introduced the notion of causal category equipped with partial involution structure as a generalized "relativity" structure. To combine this structure with "quantum" structure, which can be modelled by noncommutative algebras, especially effectively by noncommutative algebra over $\mathbb{C}$ as history has shown, we need a noncommutative algebra that reflects the structure of category: Category algebra is just the right concept.

\subsection{Category Algebra}

We introduce the notion of category algebra in this subsection, which is based on \cite{SAI}.

\begin{Definition}[Rig]
A rig $R$ is a set with two binary operations called addition and multiplication such that

\begin{enumerate}
    \item $R$ is a commutative monoid with respect to  addition with the unit $0$,
    \item $R$ is a monoid with respect to multiplication with the unit $1$,
    \item $r''(r'+r)=r''r'+r''r, \: (r''+r')r=r''r+r'r$ holds for any $r,r',r''\in R$ (Distributive law),
    \item $0r=0, \: r0=0$ holds for any $r\in R$ 
    (Absorption law).
\end{enumerate}

\end{Definition}


Note that in general a rig can be noncommutative. The notion of center is important for noncommutative rigs:

\begin{Definition}[Center]
A subrig $Z(R)$ of a rig $R$ defined as the set of elements which are commutative with all the elements in $R$ is called the center of $R$.  
\end{Definition}

A rig $R$ is commutative if and only if $Z(R)=R$.

Based on the notion of rigs, we define the notion of modules and algebras over rigs.

\begin{Definition}[Module over Rig]
A commutative monoid $M$ under addition with unit $0$ together with a left action of $R$ on $M$ $(r,m)\mapsto rm$ is called a left module over $R$ if the action satisfies the following conditions:
\begin{enumerate}
    \item $r(m'+m)=rm'+rm, \: (r'+r)m=r'm+rm$ for any $m,m'\in M$ and  $r,r'\in R$.
    \item $0m=0, \: r0=0 $ for any $m\in M$ and $r \in R$.
\end{enumerate}

Dually we can define the notion of right module over $R$.

Let $M$ is left and right module over $R$. $M$ is called an $R$-bimodule if 
\[
r'(mr)=(r'm)r
\]
holds for any $r,r'\in R$ and $m\in M$.
The left/right action above is called the scalar multiplication.
\end{Definition}


\begin{Definition}[Algebra over Rig]
A bimodule $A$ over $R$ is called an algebra over $R$ if it is also a rig with respect to its own multiplication which is compatible with scalar multiplication, i.e., 
\[
(r'a')(ar)=r'(a'a)r, \; (a'r)a=a'(ra)
\]
for any $a,a' \in A$ and $r,r'\in R$.
\end{Definition}


We define the principal notion of the present paper:

\begin{Definition}[Category Algebra]
Let $\mathcal{C}$ be a category and $R$ be a rig. 
An $R$-valued function $\alpha$ defined on $\mathcal{C}$ is said to be of finite propagation if for any object $C$ there are at most finite number of arrows whose codomain or domain is $C$ in its support.
The module over $R$ consisting of all $R$-valued functions of finite propagation together with the multiplication defined by
\[
({\alpha}' \alpha)(c'') = \sum_{\{(c',c)|\:c''=c'\circ c\}} {\alpha}'(c'){\alpha}(c), \:\: c,c',c''\in \mathcal{C}
\]
becomes an algebra over $R$ with unit $\epsilon$ defined by
\[
\epsilon (c)=  \begin{cases}
            1 & (c \in |\mathcal{C}|) \\
            0 & (otherwise),
            \end{cases}
\]
and is called the category algebra of finite propagation which is denoted as  $R[\mathcal{C}]$. In the present paper, we simply call $R[\mathcal{C}]$ the  category algebra of $\mathcal{C}$.
\end{Definition}

The multiplication defined above is nothing but "convolution" operation on the category $\mathcal{C}$. 
$R[\mathcal{C}]$ coincides with the algebra studied in \cite{MIT} if $R$ is a ring. In \cite{SAI} it is denoted as $^{0}R_{0}[\mathcal{C}]$ to distinguish them from other kinds of category algebras.

A functor from one category to another induces a homomorphism between the corresponding category algebras if the functor is bijective on objects. If the bijective-on-objects functor is also injective on arrows, the induced morphism becomes injective. Hence, the category algebra $R[\mathcal{C}^{\circ}]$ of a subcategory $\mathcal{C}^{\circ}$ of a category $\mathcal{C}$ becomes a subalgebra of $R[\mathcal{C}]$.

\begin{Definition}[Indeterminate]
Let $R[\mathcal{C}]$ be a category algebra and $c\in \mathcal{C}$. The function $\iota^{c} \in \: R[\mathcal{C}]$ defined as
\[
\iota^{c}(c')=  \begin{cases}
            1 & (c'=c) \\
            0 & (otherwise)
            \end{cases}
\]
is called the indeterminate corresponding to $c$.
\end{Definition}

In the previous work \cite{SAI} we denote the indeterminate $\iota^{c}$ as $\chi^{c}$. We change the notation to avoid confusion with "character" in representation theory.

For indeterminates, it is easy to obtain the following:

\begin{Theorem}[Calculus of Indeterminates]
Let $c,c'\in \mathcal{C}$, $\iota^{c},\iota^{c'}$ be the corresponding indeterminates and $r\in R$. Then
\[
\iota^{c'}\iota^{c}=  \begin{cases}
            \iota^{c'\circ c} & (\rm{dom}(c')=\rm{cod}(c)) \\
            0 & (otherwise),
            \end{cases}
\]
\[
r\iota^{c}=\iota^{c}r.
\]

\end{Theorem}

In short, a category algebra $R[\mathcal{C}]$ is an algebra of functions on $\mathcal{C}$ equipped with the multiplication which reflects the compositionality structure of $\mathcal{C}$. By the identification of $c\in \mathcal{C} \mapsto \iota^{c}\in \: R[\mathcal{C}]$, categories are included in category algebras.



A category algebra can be considered as a generalized matrix algebra. In fact, matrix algebras are isomorphic to category algebras of indiscrete categories. For the basic notions and rules for matrix-like calculation in category algebras, see \cite{SAI}. 

For the main application of the present paper, we need the involution structure on algebras:

\begin{Definition}[Involution on Rig]\label{involution on rig}
Let $R$ be a rig. An operation $(\cdot)^{\ast}$ on $R$ preserving addition and 
covariant (resp. contravariant) 
with respect to multiplication is said to be a 
covariant (resp. contravariant) 
involution on $R$ when $(\cdot)^{\ast} \circ (\cdot)^{\ast}$ is equal to the identity function on $R$. A rig with contravariant involution is called a $^{\ast}$-rig.
\end{Definition}

\begin{Definition}[Involution on Algebra]\label{involution of algebra}
Let $A$ be an algebra over a rig $R$ with a covariant (resp. contravariant) involution $\overline{(\cdot)}$ . A covariant (resp. contravariant) involution $(\cdot)^{\ast}$ on $A$ as a rig is said to be a covariant (resp. contravariant) involution on $A$ as an algebra over $R$ if it is compatible with scalar multiplication, i.e., 
\[
(r'ar)^{\ast}=\overline{r'}a^{\ast}\overline{r} \:\:\: \text{(covariant case)},\:\:\: 
(r'ar)^{\ast}=\overline{r}a^{\ast}\overline{r'} \:\:\: \text{(contravariant case)}. 
\]
An algebra $A$ over a $^{\ast}$-rig $R$ with contravariant involution is called a $^{\ast}$-algebra over $R$.
\end{Definition}

\begin{Theorem}[Category Algebra as Algebra with Involution]
Let $\mathcal{C}$ be a category with a covariant (resp. contravariant) involution $(\cdot)^{\dagger}$  and $R$ be a rig with a covariant (resp. contravariant) involution $\overline{(\cdot)}$. Then the category algebra $R[\mathcal{C}]$ becomes an algebra with covariant involution (resp. ${}^{\ast}$-algebra) over $R$.
\end{Theorem}

\subsection{Quantum Fields as Category Algebra}

In this section, we will show that category algebras provide  appropriate models for quantum fields. As already mentioned in the introduction, a quantum field is intuitively a synthesis of relativistic and quantum structures.
In the previous section, we argued that the relativistic structure as the basic structure of possible dynamics can be understood from a general point of view by the causal category. The next problem is to construct a noncommutative algebra which is consistent with relativistic covariance, as well as with causality. The category algebra $R[\mathcal{C}]$, where $C$ is a causal category equipped with partial involution structure, is just such an algebra.

Note that by generalizing groupoid algebras to category algebras, we can naturally incorporate processes that are not necessarily reversible. If we focus on the core of the category, i.e., the subcategory consisting of all invertible arrows, we have the corresponding groupoid algebra which is an subalgebra of $R[\mathcal{C}]$.

\begin{Definition}[Quantum Field]
Let $\mathcal{C}$ be a causal category with partial involution structure $\mathcal{C}^{\sim}$ and $R$ be a rig with involution. The category algebra $R[\mathcal{C}]$ is called the quantum field on $\mathcal{C}$ over $R$.
\end{Definition}

For quantum physics, the cases in which $R$ is some $^{\ast}$-algebra over $\mathbb{C}$ are important. The category $\mathcal{C}$ is considered as "spacetime with inner degrees of freedom of the field". Note that a quantum field on a causal category $\mathcal{C}$ over a rig $R$ might be isomorphic to or embedded into another quantum field on another causal category $\mathcal{C'}$ over another rig $R'$. Hence, even if we focus on the case that $R=\mathbb{C}$, we might cover many kinds of quantum fields. 
Nevertheless, we keep to let $R$ be general rig $R$ with involution when we can in the present paper for future applications. 

Let us see how a quantum field as a category algebra incorporates the relativistic covariance structure:
To begin with, let us assume that a group $G$ (say, the Poincar\'{e} group) acts on $|\mathcal{C}|$ and there is a map $u^{(\cdot )}$ sending a pair $(g,C)\in G\times \mathcal{C}$ to the arrow
$u^{(g,C)}:C\longrightarrow gC$ in $\mathcal{C}$ satisfying
$u^{(g'g,C)}=u^{(g',gC)}\circ u^{(g,C)}$ and $u^{(e,C)}=C$, 
where $e$ denotes the unit of $G$ and $C$ denotes the identity arrow on $C$ in the last equation. Note that each $u^{(g,C)}$ is an invertible arrow. Then we can define the endfunctor $\widetilde{{u}^{g}}:\mathcal{C}\longrightarrow \mathcal{C}$ by
\[
\widetilde{{u}^{g}}(c)=u^{(g,\mathrm{cod}(c))}\circ c \circ (u^{(g,\mathrm{dom}(c))})^{-1}
\]
which becomes invertible and induces the corresponding isomorphism on  category algebra $R[\mathcal{C}]$. Note also that $u^{(g,\cdot)}$ becomes a natural equivalence between $\mathcal{C}$ (identity functor on $\mathcal{C}$) to $\widetilde{{u}^g}$. 

In general, given a natural equivalence $u$ from the identity functor $\mathcal{C}$ to an invertible functor 
$\hat{u}$ from $\mathcal{C}$ to $\mathcal{C}$, we can define an invertible element $\iota^{u} \in R[\mathcal{C}]$ as
\[
\iota^{u} (c)= \begin{cases}
            1 & (c\: \text{is a component of $u$)} \\
            0 & (\text{otherwise}),
            \end{cases}
\]
and isomorphism $\widetilde{\iota^{u}}$ on  $R[\mathcal{C}]$ as

\[
\widetilde{\iota^{u}}(\alpha)=\iota^{u} \alpha {(\iota^{u})}^{-1}.
\]

This kind of transformation will be useful to study flows, generators, and symmetries such as local gauge invariance from the viewpoint of category algebras. To sum up, the category algebra intrinsically incorporates covariance structure coherent with the structure of "spacetime" category $\mathcal{C}$.


In order to consider the essential features of relativity, it is necessary to consider the structure of causal categories. For this purpose, let us consider the category algebras on relevant categories and relevant categories with involution.

\begin{Definition}[Relevant Algebra and Local Algebra]
Let $R$ be a rig and $\mathcal{C}$ be a causal category. 
The category algebra $R[\mathcal{O}^{rel}]$ is called the relevant algebra on $\mathcal{O}$, 
a subset of $|\mathcal{C}|$, over $R$. 
A subset $\mathcal{O}$ of $|\mathcal{C}|$ is called a region if for any $C\in |\mathcal{C}|\setminus \mathcal{O}$ there is no nontrivial factorization in $\mathcal{O}^{rel}$.
For a region $\mathcal{O}$, the subrig $R^{loc}[\mathcal{O}]$ of $R[\mathcal{O}^{rel}]$ whose elements are in the form of $\alpha + \delta$, where $\alpha$ denotes an element in $R[\mathcal{O}^{rel}]$ satisfying $\alpha (C)=0$ for any $C\in |\mathcal{C}|\setminus \mathcal{O}$ and $\delta$ denotes an element in $R[\mathcal{O}^{rel}] \cap Z(R[\mathcal{C}])$, becomes an algebra over $Z(R)$ and called the local algebra on $\mathcal{O}$.
\end{Definition}

\begin{Definition}[Relevant Algebra/Local Algebra with Involution]
Let $R$ be a rig with involution and $\mathcal{C}$ be a causal category with partial involution structure. 
The category algebra $R[\mathcal{O}^{rel\sim}]$ is called the relevant algebra with involution on $\mathcal{O}$, a subset of $|\mathcal{C}|$, over $R$. For a region $\mathcal{O}$, the subrig $R^{loc\sim}[\mathcal{O}]$ of $R[\mathcal{O}^{rel\sim}]$ whose elements are in the form of $\alpha + \delta$, where $\alpha$ denotes an element in $R[\mathcal{O}^{rel\sim}]$ satisfying $\alpha (C)=0$ for any $C\in |\mathcal{C}|\setminus \mathcal{O}$ and $\delta$ denotes an element in $R[\mathcal{O}^{rel\sim}] \cap Z(R[\mathcal{C}])$, becomes an algebra with involution over $Z(R)$ and called the local algebra with involution on $\mathcal{O}$.

\end{Definition}



The family of local algebras with involution $\{R^{loc}[\mathcal{O}]\}$, especially when $R$ is a $^{\ast}$-algebra over $\mathcal{C}$, is the counterpart of $\{\mathcal{A}(\mathcal{O})\}$ in AQFT \cite{HAG}, where $\mathcal{A}(\mathcal{O})$ denotes the observable algebra defined on bounded region $\mathcal{O}$ in the spacetime.
Our framework so far does not focus on the topological aspect of algebras, but the conceptual correspondence between our framework and AQFT is remarkable as we will see below.

Note that our "local" algebras in general contain certain kind of information of "outside" of the regions. Nevertheless, they contain no information of the local algebras corresponding to spacelike separated regions.
From the structure theorem of relevant category and the definition of local algebras, we have the following:

\begin{Theorem}[Commutativity of Spacelike Separated Local Algebras] 
Local algebras $R^{loc}[\mathcal{O}]$ and 
$R^{loc}[\mathcal{O'}]$ are commutative with each other if the regions $\mathcal{O}$ and $\mathcal{O'}$ are spacelike separated each other.
\end{Theorem}

As a collorary we have:

\begin{Theorem}[Commutativity of Spacelike Separated Local Algebras with Involution] 
Local algebras $R^{loc\sim}[\mathcal{O}]$ and 
$R^{loc\sim}[\mathcal{O'}]$ with involution are commutative with each other if the regions $\mathcal{O}$ and $\mathcal{O'}$ are spacelike separated from each other.
\end{Theorem}

The theorem above is the conceptual counterpart of one of the axiom called "Einstein causality" ("Axiom E" in \cite{HAG}).

\subsection{Remarks on  Comparison to TQFT}

Our category algebraic framework of quantum field theory can also be compared to the conceptual ideas in other axiomatic approaches to quantum fields such as Topological Quantum Field Theory (TQFT) \cite{ATI,WIT}.
In the axiomatization of TQFT, a quantum field theory is considered as a certain functor from the category of $n$-cobordism $nCob$ into the category $Mod(R)$ of modules over some unital commutative ring $R$ (the typical case is $R=\mathbb{C}$ and $Mod(R)=Vect$, where $Vect$ denotes the category of vector spaces over $\mathbb{C}$).

We can construct such a functor in a generalized setting based on our framework: 
Let $C$ be an object in a $^{\dagger}$-category $\mathcal{C}$ and $R$ be a rig. We define the submodule  $^{C}R[\mathcal{C}]$ of $R[\mathcal{C}]$ consisting of elements whose support is included in the set of arrows whose codomain is $C$. 
Then we can define a functor $^{(\cdot)}R:\mathcal{C}\longrightarrow Mod(R)$ by $^{c}R=\; \iota^{c}(\cdot)$, the multiplication of $\iota^{c}$, i.e. a module homomorphism sending each $\alpha \in{}^{\mathrm{dom}(c)}R[\mathcal{C}]$ to $\iota^{c}\alpha \in{}^{\mathrm{cod}(c)}R[\mathcal{C}]$, for any $c\in \mathcal{C}$.

Since $nCob$ is a $\dagger$-category, the above construction works and we obtain a canonical functor from $nCob$ to $Mod(R)$. Although this functor does not satisfy all the technical parts of axioms proposed in TQFT, 
it is coherent with the physical ideas of relativistic covariance and quantum properties behind the axioms. This coherence will become more clear after introducing the notion of states on categories in the next section.

\section{States of Quantum Fields as States on Categories}

\subsection{State on Category}

While an algebra embodies the intrinsic structure of a system, a state embodies the interface between that system and its environment. This view, which has been advocated by Ojima \cite{OJIjapanese}, is consistent with the mathematical framework of algebraic quantum field theory and quantum probability theory: states provide concrete representations of an algebra. 

In general, a representation refers to an expression of intrinsic structures in a certain way, which corresponds to the concrete realization of the intrinsic properties of a system in the way it interacts with its environment.
To be more specific, a state is a mapping which sends elements of an algebra to a scalar values as "expectation values". In short, states define the statistical laws, which generalize the notion of probability measures to the noncommutative context. Conversely, if the algebra is a unital commutative $C^{\ast}$-algebra we have Radon measure on a compact Hausdorff space by Riesz-Markov-Kakutani Theorem \cite{HO}. In other words, a pair of an algebra and state on it is a generalized probability space: a noncommutative probability space.

As for the category algebras that reflect the structure of the possible dynamics, defining a state on it means evaluating  arrows corresponding to the individual processes with expectation values. Conversely, for a category with a finite number of objects, the weighting of the arrows gives a state. Based on this fact, we call a state on a category algebra a state on category by abuse of terminology.  

The rest of this subsection is based on \cite{SAI}.

\begin{Definition}[Linear Functional]
Let $A$ be an algebra over a rig $R$. An $R$-valued linear function on $A$, i.e., a function preserving addition and scalar multiplication, is called a linear functional on $A$.
A linear functional on $A$ is said to be unital if  $\varphi(\epsilon)=1$ where $\epsilon$ and $1$ denote the multiplicative unit in $A$ and $R$, respectively.
\end{Definition}


\begin{Definition}[Positivity]\label{positivity}
A pair of rigs with involution $(R,R_{+})$ is called a positivity structure on $R$ if $R_{+}$ is a subring with involution such that $r,s \in R_{+}$ and $r+s=0$
imply $r=s=0$, and that $a^{\ast}a\in R_{+}$ for any $a\in R$. 

\end{Definition}

\begin{Definition}[State]\label{state}
Let $R$ be a rig with involution 
and $(R,R_{+})$ be a positivity structure on $R$. A state $\varphi$ on an algebra $A$ with involution 
over $R$ with respect to $(R,R_{+})$ is a unital linear functional $\varphi :  A\longrightarrow R$ which satisfies $\varphi(a^{\ast}a) \in R_{+}$ and 
$\varphi(a^{\ast})=\overline{\varphi(a)}$ for any $a\in R$, where $(\cdot)^{\ast}$ and $\overline{(\cdot)}$ denotes the  involution on $A$ and $R$, respectively. (The last condition $\varphi(a^{\ast})=\overline{\varphi(a)}$ follows from other conditions, if $R=\mathbb{C}$.)
\end{Definition}

\begin{Definition}[Noncommutative Probability Space]
A pair $(A,\varphi)$ consisting of an algebra $A$ with involution over a rig $R$ with involution and a state $\varphi$ is called a noncommutative probability space.
\end{Definition}

\begin{Definition}[State on Category]

Let $R$ be a rig with involution, $(R,R_{+})$ be a positivity structure on $R$, and $\mathcal{C}$ be a category with involution. A state on the category algebra $R[\mathcal{C}]$ over $R$ with respect to $(R,R_{+})$ is said to be a state on a category $\mathcal{C}$ with respect to $(R,R_{+})$.
\end{Definition}

Given a state $\varphi$ on a category $\mathcal{C}$ with involution, we have an function $\hat{\varphi}:\mathcal{C}\longrightarrow R$ defined as $\hat{\varphi}(c)=\varphi(\iota^{c})$. For the category with finite number of objects, we can obtain the following theorem \cite{SAI}, which is a generalization of the result in \cite{CIM} for groupoids:

\begin{Theorem}[State and Normalized Positive Semidefinite Function]
Let $\mathcal{C}$ be a category with involution such that $|\mathcal{C}|$ is finite. Then there is a one-to-one correspondence between states $\varphi$ with respect to $(R,R_{+})$ and normalized positive semidefinite $Z(R)$-valued functions $\phi$ with respect to $(R,R_{+})$, i.e., normalized functions such that
\[
\sum_{\{(c,c')
| \rm{dom}((c')^{\dagger})=\rm{cod}(c)\}}\overline{\xi(c')}\phi((c')^{\dagger}\circ c)\xi(c)
\]
is in $R_{+}$ for any $R$-valued function $\xi$ on $\mathcal{C}$ with finite support and that $\phi(c^{\dagger})=\overline{\phi(c)}$, where $(\cdot)^{\ast}$ and $\overline{(\cdot)}$ denotes the  involution on $\mathcal{C}$ and $R$, respectively.
\end{Theorem}

Conceptually, the theorem above means that states on a category with involution (with finite objects) are nothing but the weights on arrows, which are generalization of probability distribution on a (finite) set as the discrete category (with finite objects). More generally, we can say that to define a state on a category whose support is contained in a subcategory with finite numbers of objects is equivalent to define the corresponding function which assign the weight to each arrow. 

For a state on a category whose support is not contained in a subcategory with finite number of objects, we will need some topological structures (or coarse geometric structures \cite{ROE}). Nonstandard-analytical methods (see \cite{SN}, for example) will provide useful tools.

\subsection{States of Quantum Fields as States on Categories}

As we see quantum fields as category algebras, it is quite natural to model physical states of quantum fields as states on category algebras.

\begin{Definition}[State of Quantum Field]
Let $\mathcal{C}$ be a causal category with partial involution structure $\mathcal{C}^{\sim}$
and $(R,R_{+})$ be a positivity structure on a rig $R$ with involution. A unital linear functional on $\mathcal{C}$ which is also a state on $\mathcal{C^{\sim}}$ with respect to $(R,R_{+})$ whose image is contained in a subrig $R'$ with involution of $R$ is said to be an $R'$-valued state on the quantum field on $\mathcal{C}$ over $R$ with respect to $(R,R_{+})$.
\end{Definition}

In conventional cases $R$ is supposed to be a $^{\ast}$-algebra over $\mathbb{C}$ and $R'=\mathbb{C}$.
The phrase "with respect to $(R,R_{+})$" will be omitted if it is clear in the context.
In general, given a state $\varphi$ on $^{\ast}$-algebra $A$ over a $^{\ast}$-rig $R$ with involution, we can construct the representation of the algebra into the algebra consisting of endomorphism on certain module (generalized GNS representation \cite{SAI}). Especially, when $R$ is a $^{\ast}$-rig over $\mathbb{C}$ and $\varphi$ is $\mathbb{C}$-valued, we have a  representation called GNS (Gelfand-Naimark-Segal) representation \cite{GN,SEG} into a pre-Hilbert space consisting of equivalence class of the elements in $A$, equipped with the inner product structure induced by sesquilinear form $\langle a', a \rangle =\varphi ((a')^{\ast}a)$ ($a,a'\in A$). 
The unit of the algebra plays a role of "vacuum" vector (See \cite{SAI} and reference therein, for example). 

To sum up, a noncommutative probability space, i.e., a pair $(A,\varphi)$ of $^{\ast}$-algebra over $\mathbb{C}$ and $\mathbb{C}$-valued state on it, is sufficient to reconstruct the ingredients in conventional quantum physics based on Hilbert space. In fact, the approach based on the noncommutative probability space is more general than the conventional approach: if we focus into the local structures of quantum fields, it is known that we cannot use one a-priori Hilbert space as starting point of the theory (Actually, this fact itself was one of the historical motivation of AQFT, the pioneer of noncommutative probabilistic approach. See \cite{HAG} for example) . 

Our category algberaic approach is a new unification of the noncommutative probabilistic approach and the category theoretic viewpoint. As we see in the previous section, for $^{\dagger}$-category (or in general, category with involution), its category algebra is a $^{\ast}$-algebra (or in general, an algebra with involution). Note that our algebra is unital, even if $\mathcal{C}$ has infinitely many objects. Then the construction above holds and we can see the unit $\epsilon$ as "vacuum" in our theory.

The concept of states on categories also shed light on the foundation of quantum mechanics as a part of quantum field theory. 
From our viewpoint, a quantum mechanical system of finite degrees of freedom can be defined as a noncommutative probability space whose algebra is a subalgebra of a category algebra on a causal category with partial involution and whose state satisfies the condition that the support is contained in a subcategory with finite numbers of objects. 

In general, quantum fields as category algebras together with states "contains" vast numbers of quantum mechanical systems. Or more precisely, considering a state whose support is contained in some subcategory with finite numbers of objects is focusing on a quantum mechanical system as an aspect of the quantum field. 
Note that quantum mechanical systems in the above meaning  
are not necessarily contained in a single point but can have spatial degrees of freedom, e.g., a system in double slit experiment, where the support of the state can be considered to be contained in a subcategory with finite objects. 
Understanding the situation where multiple observers are involved in a single quantum field---such as the EPR (Einstein-Podolsky-Rosen) situation \cite{EPR}--- through the concepts of local algebra and local states seem also to be important research topic.

The idea at the heart of the above discussions is that we are free to think of ”localized states” (not just "global" states like vacuum states). The concept corresponding to these kinds of states is particularly important in the context of AQFT and is called "local states"\cite{WER,OOS}. We can define the local states in our framework, which is a conceptual counterpart of local states in AQFR, as follows:

\begin{Definition}[Local State]
Let $\mathcal{C}$ be a causal category equipped with partial involution structure  $\mathcal{C}^{\sim}$ and $R$ be a rig with involution. A state on $R[\mathcal{C}^{\sim}(\mathcal{O})]$ for a region $\mathcal{O}$ is called a local state of the quantum field $R[\mathcal{C}]$ on $\mathcal{O}$.
\end{Definition}

From a physical point of view, the notion of local state is quite natural. A macroscopic setting of the environment for the quantum field basically concerns only the bounded spacetime domain, and the global state should be seen as an idealization of it. Considering a family of local states instead of a single state can be seen as a sheaf theoretic extension of conventional quantum field theory. The extension will lead to the notion of consistent families of Hilbert spaces and operators on them, which will be mathematically interesting, through the GNS construction.

By translating the previous study \cite{OOS} into our context, we will be able to construct generalized sector theory which is the generalization of DHR(Doplicher-Haag-Roberts)-DR(Doplicher-Roberts) Theory\cite{DHR1,DHR2,DHR3,DHR4,DR1,DR2,DR3} and develop Ojima's micro-macro duality \cite{OJI1,OJI2} and quadrality scheme \cite{OJI3} from the viewpoint of category algebras and states on categories.

\subsection{Remarks on Comparison to TQFT (continued)}

In section 3, we have constructed for any $^{\dagger}$-category $\mathcal{C}$ a functor $^{(\cdot)}R:\mathcal{C}\longrightarrow Mod(R)$ by $^{c}R=\; \iota^{c}(\cdot)$ for $c\in\mathcal{C}$. For quantum physical studies, we need to induce a functor into $Hilb$, a category of Hilbert spaces over $\mathbb{C}$. 
Let us explain the role of states in this induction.

Given any state $\varphi$ on the $^{\dagger}$-category $\mathcal{C}$, $^{C}R:={}^{C}R[\mathcal{C}]$ for each object $C\in\mathcal{C}$ can be equipped with "almost inner product" (semi-Hilbert space structure) by defining sesquilinear form 
$\langle \cdot | \cdot \rangle^{\varphi}$
by $\langle \alpha' | \alpha \rangle^{\varphi}:=\varphi((\alpha')^{\ast}\alpha)$ (generalized GNS consruction, see \cite{SAI} and references therein). When $R$ is a $^{\ast}$-algebra over $\mathbb{C}$ and $\varphi$ is a $\mathbb{C}$-valued "good" state $\varphi$ on the category given, this functor induces a functor into $Hilb$.
More precisely, suppose that $R$ is a $^{\ast}$-algebra over $\mathbb{C}$ and the $\mathbb{C}$-valued state $\varphi$ satisfies the condition
\[
\varphi(\alpha^{\ast}\alpha)\geq 
\varphi((\iota^{c}\alpha)^{\ast}(\iota^{c}\alpha))
\]
for any $\alpha \in R[\mathcal{C}]$ and $c\in \mathcal{C}$.
Then the functor $^{(\cdot)}R:\mathcal{C}\longrightarrow Mod(R)$ induces the functor $^{(\cdot)}R^{\varphi}:\mathcal{C}\longrightarrow preHilb$, where $preHilb$ denotes the category of pre-Hilbert spaces, taking the quotient of $^{C}R$ equipped with $\langle \cdot | \cdot \rangle^{\varphi}$ by $N^{\varphi}=\{\alpha\in {}^{C}R|\varphi(\alpha^{\ast}\alpha)=0\}$, which can be shown as submodule of $^{C}R$. (Note that by the assumption 
$\varphi(\alpha^{\ast}\alpha)=0 \Longrightarrow
\varphi((\iota^{c}\alpha)^{\ast}(\iota^{c}\alpha))=0$
holds. This kind of construction itself has certain generalization to a more general $R$ by using this condition directly). Then by the assumption of $\varphi$, $^{(c)}R^{\varphi}$ extends uniquely to the morphism in $Hilb$ by completion and we have the corresponding functor from $\mathcal{C}$ to $Hilb$ which sends $c$ to the unique bounded extension of $^{(c)}R^{\varphi}$ (Note that a bounded operator between pre-Hilbert spaces extends to the unique bounded operator between Hilbert spaces). 
By applying this construction for $\mathcal{C}=nCob$, we have a version of TQFT. 
Note that our framework is naturally incorporated with causal structure and it is quite interesting to the counterpart of the structure in TQFT.

\section{Prospects}


As we have seen, our new approach to quantum fields is conceptually related to conventional approaches such as AQFT and TQFT. Elucidating this relationship at a deeper level will be important in the study of quantum fields. In order to carry out such research, we will need to include more detailed structures such as topological or differential structures in addition to the algebraic and noncommutative probabilistic structures that we have discussed in this paper.

On the other hand, it should be emphasized that our approach is directly applicable to lattice gauge theory \cite{WIL} and other discrete spacetime approaches, as can be seen from the fact that our approach deals with general categories. Its applicability extends to the context of unifying general relativity and quantum theory.

Needless to say, the relationship with categorical approaches to quantum theory such as "categorical quantum mechanics" \cite{AC1,AC2} based on the $^{\dagger}$-category should also be explored. The categorical structure of the submodules of the category algebra as a generalized matrix algebra and the computations based on it will play an important role. It is also interesting to clarify the relationship between our framework and the approach in a recently published article \cite{GSC} which also deals with AQFT and Quantum Cellular Automata (QCA) approach \cite{DP, ARR} from a general categorical viewpoint.

The notion of quantum walk (see \cite{AMB,KON} and references therein, for example), which is closely related to the QCA approach, can also be formulated from our standpoint. Based on our framework, we can model a concrete dynamics of quantum fields as a sequence or flow of the states on a category. In general, the dynamics can be irreversible. The typical examples of reversible dynamics are called quantum walks. The notion of quantum walks on general $^{\ast}$-algebras and quantum walks on $^{\dagger}$-categories can be defined as follows:

\begin{Definition}[Quantum Walk]
Let $A$ be a $^{\ast}$-algebra. A sequence of states given by 
\[
\varphi^{t}(\alpha)=\varphi((\omega^{\ast})^{t}\alpha \omega^{t})\;\;t=0,1,2,3,...
\]
generated by a unitary element $\omega \in R[\mathcal{C}]$, i.e. an element satisfying ${\omega}^{\ast}\omega=\omega \omega^{\ast}=\epsilon$ is called a quantum walk on $A$.
\end{Definition}

\begin{Definition}[Quantum Walk on $^{\dagger}$-Category]
Let $\mathcal{C}$ be a $^{\dagger}$-category and $R$ be a $^{\ast}$-rig. A quantum walk on $R[\mathcal{C}]$ is said to be a quantum walk on a $^{\dagger}$-category $\mathcal{C}$.
\end{Definition}

A quantum walk can be considered as a sequence of "state vector" through GNS construction. 
The notion of quantum walk defined on $^{\dagger}$-category includes the various concrete dynamical models under the name of quantum walks. For example, quantum walks on simple undirected graphs as a certain sequence of state on an indiscrete category. 
The category algebraic approach will play a fundamental role for the quantum walks on graphs with multiple edges and loops. Quantum walks on graphs have been used in the modeling of "dressed photon" \cite{OHT} which cannot be understood without focusing on off-shell nature of quantum fields \cite{HS}, i.e., the aspects of quantum fields cannot be described as the collection of the modes which satisfies the on-shell condition,  and quantum walks on categories may become important in quantum field theory in general. They will also connects the QCA approach to quantum fields and other approaches to quantum fields.














One of the most exciting problems is, of course, to construct a model of a non-trivial quantum field with interactions. We believe we can approach such problems. In particular, it seems that the fact that relevant categories have arrows that go through objects in very distant regions, while the local algebras defined on them satisfy commutativity, may be the key to avoiding various no-go theorems. 
Note also that our approach extends the coefficients to a general (commutative or noncommutative) rigs, which greatly expands the possibilities of dealing with interactions.
Finally, it should be pointed out that our approach is not limited to quantum fields, but can be extended to give a very general noncommutative statistical model with a causal structure. 
The author hopes that the present paper will be a small new step towards these big problems.

\section*{Acknowledgements}
The author is grateful to Prof. Hiroshi Ando, Prof. Takahiro Hasebe, Dr. Soichiro Fujii, Prof. Izumi Ojima, Dr. Kazuya Okamura, Ms. Misa Saigo and Mr. Juzo Nohmi for fruitful discussions and comments.
This work  was partially supported by Research Origin for Dressed
Photon, JSPS KAKENHI (grant number 19K03608 and 20H00001) and JST CREST (JPMJCR17N2).

\end{document}